%% file: 00_main.tex
\documentclass[reprint,superscriptaddress,nofootinbib,amsmath,amssymb,aps]{revtex4-1}

\usepackage{graphicx}
\usepackage{dcolumn}
\usepackage{bm}

\usepackage{float}
\usepackage{url}
\usepackage{hyperref}
\usepackage{cleveref}
\usepackage{braket}
\usepackage{subcaption}

\crefname{figure}{Fig.}{Figs.}
\crefname{equation}{Eq.}{Eqs.}
\crefname{section}{Sec.}{Sec.}
\Crefname{figure}{Figure}{Figures}
\Crefname{equation}{Equation}{Equations}
\Crefname{section}{Section}{Sections}

\newcommand{\loss}{\mathcal{L}}
\newcommand{\expect}[1]{\ensuremath{\langle {#1} \rangle}}

\begin{document}

\preprint{APS/123-QED}

\title{Subspace-search variational quantum eigensolver for excited states}

\author{Ken M Nakanishi}
    \email{ken-nakanishi@g.ecc.u-tokyo.ac.jp}
    \affiliation{
        Graduate School of Science,
        The University of Tokyo,
        7-3-1 Hongo, Bunkyo-ku, Tokyo 113-0033, Japan.
    }
\author{Kosuke Mitarai}
    \email{mitarai@qc.ee.es.osaka-u.ac.jp}
    \affiliation{
        Graduate School of Engineering Science,
        Osaka University,
        1-3 Machikaneyama, Toyonaka, Osaka 560-8531, Japan.
    }
    \affiliation{
        QunaSys Inc.,
        High-tech Hongo Building 1F, 5-25-18 Hongo, Bunkyo, Tokyo 113-0033, Japan.
    }
\author{Keisuke Fujii}
    \email{fujii.keisuke.2s@kyoto-u.ac.jp}
    \affiliation{
        Graduate School of Science,
        Kyoto University,
        Kitashirakawa Oiwake-cho, Sakyo-ku, Kyoto 606-8302, Japan.
    }
    \affiliation{
        JST, PRESTO, 4-1-8 Honcho, Kawaguchi, Saitama 332-0012, Japan.
    }

\date{\today}

\input{01_abstract.tex}

\pacs{Valid PACS appear here}
\maketitle

\input{10_intro.tex}
\input{30_method.tex}
\input{31_SSVQE.tex}
\input{33_WSSVQE_k.tex}
\input{35_WSSVQE_up_to_k.tex}
\input{40_relatedworks.tex}
\input{50_application.tex}
\input{60_experiment.tex}
\input{80_conclusion.tex}
\input{85_ack.tex}

\vspace{10mm}

\bibliography{bibliography}

\end{document}

%% file: 01_abstract.tex
\begin{abstract}
The variational quantum eigensolver (VQE), a variational algorithm to obtain an approximated ground state of a given Hamiltonian, is an appealing application of near-term quantum computers.
The original work [A. Peruzzo et al.; \textit{Nat. Commun.}; \textbf{5}, 4213 (2014)] focused only on finding a ground state, whereas the excited states can also induce interesting phenomena in molecules and materials.
Calculating excited states is, in general, a more difficult task than finding ground states for classical computers.
To extend the framework to excited states, we here propose an algorithm, the subspace-search variational quantum eigensolver (SSVQE).
This algorithm searches a low energy subspace by supplying orthogonal input states to the variational ansatz and relies on the unitarity of transformations to ensure the orthogonality of output states.
The $k$-th excited state is obtained as the highest energy state in the low energy subspace.
The proposed algorithm consists only of two parameter optimization procedures and does not employ any ancilla qubits.
The disuse of the ancilla qubits is a great improvement from the existing proposals for excited states, which have utilized the swap test, making our proposal a truly near-term quantum algorithm.
We further generalize the SSVQE to obtain all excited states up to the $k$-th by only a single optimization procedure.
From numerical simulations, we verify the proposed algorithms.
This work greatly extends the applicable domain of the VQE to excited states and their related properties like a transition amplitude without sacrificing any feasibility of it. 
\end{abstract}

%% file: 10_intro.tex
\section{Introduction}

Supported by the world-wide active research for the development of quantum devices, quantum computers equipped with almost a hundred qubits are now within reach.
Those near-term quantum computers are often called noisy intermediate-scale quantum (NISQ) devices~\cite{Preskill2018}, reflecting the fact that those quantum computers are not fault-tolerant, that is, they do not have the guaranteed accuracy of the computational result.
However, such a NISQ device is believed not to be simulatable on classical computers if the gate fidelity is sufficiently high~\cite{Boixo2018,Bouland2018,Chen2018}.
This fact encourages us to look for practical applications of them.

The variational quantum eigensolver (VQE)~\cite{Peruzzo2014,Bauer2016,Kandala2017} is an attracting application of near-term quantum computers in the hope that the controllable quantum devices can simulate another quantum system more efficiently than classical devices.
The VQE is an algorithm for finding an approximate ground state of a given Hamiltonian $H$.
For this purpose, the VQE utilizes a parameterized quantum circuit $U(\bm{\theta})$, which is also called an ansatz circuit, to generate an ansatz state $\ket{\psi(\bm{\theta})}$.
The expectation value of the target Hamiltonian $\expect{H (\bm{\theta})} = \bra{\psi(\bm{\theta})}H\ket{\psi(\bm{\theta})}$ is minimized by iterative optimization of the parameters $\bm{\theta}$.
The circuit with the resultant optimal parameters $\bm{\theta}^*$ which minimizes $\expect{H}$ outputs the approximate ground state.

Not only the ground state, which the original VQE aims to find, but also excited states of molecules are responsible for many chemical reactions and physical processes.
For example, the transition between a ground state and excited states is the origin of luminescence~\cite{Klessinger1995}.
Intermediate states of a chemical reaction are, in general, not a ground state of a system, and therefore properties of such excited states are important for the analysis of them~\cite{Lischka2018}.

In spite of the importance of the excited states, classical computation suffers from the increasing computational cost and gives relatively poor results for them~\cite{Serrano-Andres2005,Dreuw2005,Lischka2018}.
This motivates us to utilize quantum computers for the task of finding excited states and analyzing their property.
A long-term quantum algorithm for a chemical reaction has been investigated in Ref.~\cite{Reiher2017}.
However, algorithms which we can run on NISQ devices are yet to appear.

In order to find the excited states using NISQ devices, we propose a method, which utilizes the conservation of orthogonality under the unitary transformation.
We name the method as the subspace-search VQE (SSVQE).
The SSVQE takes two or more orthogonal states as inputs to a parametrized quantum circuit, and minimizes the expectation value of the energy in the space spanned by those states.
This method automatically imposes the orthogonality condition on the output states, and therefore allows us to remove the swap test~\cite{Buhrman2001}, which has been employed in the previous works~\cite{Higgott2018,Endo2018} to ensure the orthogonality.
In principle, the proposed algorithm can find the $k$-th excited state by running optimization of the circuit parameters only twice.
We also propose a generalized version of the SSVQE, which finds all excited states up to the $k$-th only one optimization procedure.
As a possible application of the SSVQE, a method to measure a transition amplitude between two eigenstates is described.
It can evaluate material properties such as permittivity and rate of spontaneous emission.
We perform numerical simulations and show validity of proposed algorithms for fully connected random transverse Ising models and Helium hydride.
This work greatly extends the practicability of the VQE by enabling it to find the excited states efficiently, and thereby, pushes the VQE as a candidate for a possible application of NISQ devices further.

The rest of the paper is organized as follows.
In \cref{sec:methods} we first propose the algorithm of the SSVQE and the extended version of it.
Then, in \cref{sec:relatedworks} we briefly review the existing works addressing the same objective of finding the excited states in the framework of the VQE.
An algorithm to obtain the transition amplitude is described in \cref{sec:application}.
Finally, we present the simple, proof-of-principle numerical simulations in \cref{sec:numerical_simulation}.

%% file: 30_method.tex
\section{Methods}\label{sec:methods}

The VQE is a quantum-classical hybrid algorithm to find a ground state of a given Hamiltonian $H$ using NISQ devices.
For this purpose, the VQE utilizes a parameterized quantum circuit $U(\bm{\theta})$, also called an ansatz circuit, to generate an ansatz state $\ket{\psi(\bm{\theta})}$.
The expectation value of the target Hamiltonian $\expect{H (\bm{\theta})} = \bra{\psi(\bm{\theta})}H\ket{\psi(\bm{\theta})}$ is minimized by iterative optimization of the parameters $\bm{\theta}$.

Our objective here is to find excited states of the Hamiltonian $H$.
Since the eigenstates of the Hamiltonian $H$ are mutually orthogonal, a straightforward construction of the algorithm to find $k$-th excited state is to minimize $\expect{H(\bm\theta)}$ imposing an orthogonality condition between the ansatz state $\ket{\psi(\bm{\theta})}$ and all of the ground/excited states up to $(k-1)$-th.
By inductively repeating this, we can find the excited states of interest.
The swap test~\cite{Buhrman2001}, which can measure the inner product between the ground state and the ansatz state, has been employed to ensure the orthogonality in the previous works~\cite{Higgott2018,Endo2018}.
In contrast, the SSVQE and the weighted SSVQE we propose here utilize the conservation of orthogonality under the unitary transformation in an effort to satisfy the orthogonality condition.
These methods automatically impose the orthogonality on the output states, and therefore remove the swap test.

%% file: 31_SSVQE.tex
\subsection{Subspace-search variational quantum eigensolver}
The key idea is to ensure the orthogonality at the \textit{input} of the quantum circuit, not at the output.
Below we describe the algorithm to find the $k$-th excited state that works on an $n$-qubit quantum computer.
We define the ground state as the $0$-th excited state.
The algorithm, which we refer to as the subspace-search VQE (SSVQE), runs as follows.

\textbf{Algorithm:}
\begin{enumerate}
    \item Construct an ansatz circuit $U(\bm{\theta})$ and choose input states $\left\{\ket{\varphi_j}\right\}_{j=0}^k$ which are mutually orthogonal ($\braket{\varphi_i|\varphi_j}=\delta_{ij}$).
    \item Minimize $\loss_1(\bm{\theta}) = \sum_{j=0}^k \braket{\varphi_j|U^\dagger(\bm{\theta})HU(\bm{\theta})|\varphi_j}$. We denote the optimal $\bm{\theta}$ by $\bm{\theta}^*$.
    \item Construct another parametrized quantum circuit $V(\bm{\phi})$ that only acts on the space spanned by $\left\{\ket{\varphi_j}\right\}_{j=0}^k$.
    \item Choose an arbitrary index $s\in\{0,\cdots,k\}$, and maximize 
    $\loss_2(\bm{\phi}) =
    \braket{\varphi_s|V^\dagger(\bm{\phi})U^\dagger(\bm{\theta}^*)HU(\bm{\theta}^*)V(\bm{\phi})|\varphi_s}.$
\end{enumerate}
We note that, in practice, the input states $\left\{\ket{\varphi_j}\right\}_{j=0}^k$ will be chosen from a set of states which are easily preparable, such as the computational basis.

Let the set of eigenstates of $H$ be $\left\{\ket{E_j}\right\}_{j=0}^{2^n -1}$ with corresponding eigenenergies $\left\{E_j\right\}_{j=0}^{2^n -1}$ where $E_i\geq E_j$ when $i\geq j$.
Then, the circuit optimized by the step 2 of the above algorithm is a unitary that best approximates the mapping from the space spanned by $\left\{\ket{\varphi_j}\right\}_{j=0}^k$ to one spanned by $\left\{\ket{E_j}\right\}_{j=0}^k$.
Therefore, in step 2, we can find the subspace which includes $\ket{E_k}$ as the highest energy state, using a carefully constructed ansatz $U(\bm{\theta})$.
The unitary $V(\bm{\phi})$ is responsible for searching in that subspace.
By maximizing $\loss_2(\bm{\phi})$, we find the $k$-th excited state $\ket{E_k}$.

In the case of $k \geq 2^{n-1}$, it is faster to choose $2^n-k$ of orthogonal input states $\ket{\varphi_j}$ and maximize $\loss_1(\bm{\theta})$ instead of minimizing it in the step 2, then minimize $\loss_2(\bm{\theta})$ instead of maximizing it in the final step.

%% file: 33_WSSVQE_k.tex
\subsection{Weighted SSVQE for finding the $k$-th excited state}\label{sec:kth-WSSVQE}
Here we extend the algorithm described in the previous section to find the $k$-th excited state of a given Hamiltonian which requires only a single optimization procedure.
It runs as follows.

\textbf{Algorithm:}
\begin{enumerate}
    \item Construct an ansatz circuit $U(\bm{\theta})$ and choose input states $\left\{\ket{\varphi_j}\right\}_{j=0}^k$ which are orthogonal with each other ($\braket{\varphi_i|\varphi_j}=\delta_{ij}$).
    \item Minimize 
    $\loss_{w}(\bm{\theta}) = 
    w \braket{\varphi_k|U^\dagger(\bm{\theta})HU(\bm{\theta})|\varphi_k} +
    \sum_{j=0}^{k-1} \braket{\varphi_j|U^\dagger(\bm{\theta})HU(\bm{\theta})|\varphi_j}$
    , where the weight $w$ can be any value in $(0, 1)$.
\end{enumerate}

When the cost $\loss_{w}$ reaches its global optimum, the circuit $U(\bm{\theta})$  becomes a unitary which maps $\ket{\varphi_k}$ to the $k$-th excited state $\ket{E_k}$ of the Hamiltonian and others to the subspace spanned by $\left\{\ket{E_j}\right\}_{j=0}^{j=k-1}$.
Therefore, by minimizing the cost $\loss_{w}$, we can find the $k$-th excited state by a single optimization process.
Note that the overall time required for the optimization might increase, due to the more complicated landscape of the cost function.

%% file: 35_WSSVQE_up_to_k.tex
\subsection{Weighted SSVQE for finding up to the $k$-th excited states}\label{sec:WSSVQE}
We further generalize the above argument and propose an algorithm for finding all excited states of a given Hamiltonian up to the $k$-th with only one optimization procedure.

\textbf{Algorithm:}
\begin{enumerate}
    \item Construct an ansatz circuit $U(\bm{\theta})$ and choose input states $\left\{\ket{\varphi_j}\right\}_{j=0}^k$ which are orthogonal with each other ($\braket{\varphi_i|\varphi_j}=\delta_{ij}$).
    \item Minimize $\loss_{\bm{w}}(\bm{\theta}) = \sum_{j=0}^k w_j \braket{\varphi_j|U^\dagger(\bm{\theta})HU(\bm{\theta})|\varphi_j}$, where the weight vector $\bm{w}$ is chosen such that $w_i > w_j$ when $i < j$
\end{enumerate}

The weight vector introduced here has the effect of choosing which $\ket{\varphi_j}$ is converted to which excited state.
It is easy to see the circuit $U(\bm{\theta})$ when the cost $\loss_{\bm{w}}$ reaches its global optimum becomes a unitary which maps $\ket{\varphi_j}$ to the $j$-th excited state $\ket{E_j}$ of the Hamiltonian for each $j \in \{0, 1, \cdots k \}$.
In this case, too, note that the overall time required for the optimization might increase due to the same reason as the previous section.

%% file: 40_relatedworks.tex
\section{Related works}\label{sec:relatedworks}

In this section, we first overview previous works, and then point out the advantages of our methods over them.

Ref.~\cite{Santagati2018} has proposed a method which hybridizes the quantum phase estimation algorithm and the VQE.
Although it is experimentally demonstrated~\cite{Santagati2018}, the method is unlikely to be implemented on a NISQ device, due to the need for the controlled time evolution.

In Ref.~\cite{Colless2018}, a method called quantum subspace expansion has been proposed.
The algorithm first finds the ground state $\ket{E_0}$ by the usual VQE protocol, and then measures the matrix elements of the Hamiltonian with respect to the space spanned by $\left\{O_\alpha\ket{E_0}\right\}$, where $\{O_\alpha\}$ is a set of excitation operators.
The diagonalization of the matrix, which is done classically, can determine the approximate eigenvalue spectra.
They have used a set of one electron excitation operators $\{a_v^\dagger a_o\}$, where $a_i^\dagger$ and $a_j$ are the fermion creation and annihilation operators respectively and $i,j$ running all possible indices, for $\{O_\alpha\}$.

The constrained VQE proposed in Ref.~\cite{Ryabinkin2018} can also be used for finding a certain set of excited states.
They proposed a way to introduce constraints, such as the number of electrons or the overall spin of the system, on the VQE.
The introduction of the constraints is done by adding the penalty term to the cost function.
Their method finds the lowest energy state under the constraints.
Since the difference in the constraints, such as the difference in the number of electrons or the overall spins, generally changes the energy of the system, it can be utilized to find a certain set of excited states.

Ref.~\cite{Higgott2018} has recently proposed an inductive method which adds a penalty term to ensure the orthogonality of the ansatz state with respect to the low-lying state.
To be more concrete, to find the $k$-th excited state, they use $\expect{H(\bm{\theta}_k)} + \sum_{i=0}^{k-1} \beta_i \left|\braket{\psi(\bm{\theta}_k)|\psi(\bm{\theta}_i^*)}\right|^2$, where $\bm{\theta}_i^*$ is the optimal parameters for the $i$-th excited state and $\beta_i$ is a hyperparameter that determines the strength of the penalty, as the target cost function to be minimized by tuning $\bm{\theta}_k$.
To estimate the overlap, their method uses the swap test, which requires us to double the number of qubits with additional gates.  
Their method works well when the hyperparameter $\beta_i$ is set properly as shown in Ref.~\cite{Higgott2018}.
Ref.~\cite{Endo2018_2} has enabled the optimization by the imaginary time evolution of the parameters in this approach.


The advantages of our methods, when compared to the methods above, are as follows.
\begin{enumerate}
    \item The energy spectrum found by the SSVQE or the weighted SSVQE is exact when $U(\bm{\theta})$ and $V(\bm{\phi})$ have the ability to represent the exact unitary which maps the $k$ input states to the $k$ eigenstates of the Hamiltonian.
    \item The swap test is not employed and thus easily implementable on the NISQ devices.
    \item In the SSVQE, there are no hyperparameters.
    \item In the weighted SSVQE, the results are unique regardless of the value of the hyperparameters if they meet the conditions.
    \item Optimization runs only twice for the SSVQE and only once for the weighted SSVQE.
\end{enumerate}

%% file: 50_application.tex
\section{Calculation of transition matrix elements}\label{sec:application}
It is possible to measure a transition amplitude of an operator $A$, $\braket{E_i|A|E_j}$, using the result of the SSVQE.
Note that $\braket{E_i|A|E_j} = \braket{\varphi_i|U^\dagger(\bm{\theta}^*) A U(\bm{\theta}^*) |\varphi_j}$ where $U(\bm{\theta}^*)$ is the optimized unitary.
We can measure this by expanding it as:
\begin{align}
    \mathrm{Re}(\braket{\varphi_i|U^\dagger(\bm{\theta}^*) A U(\bm{\theta}^*) |\varphi_j}) &=
    \braket{+^x_{ij}|U^\dagger(\bm{\theta}^*) A U(\bm{\theta}^*) |+^x_{ij}} \nonumber\\
    &\quad - \frac{1}{2}\braket{\varphi_i|U^\dagger(\bm{\theta}^*) A U(\bm{\theta}^*) |\varphi_i} \nonumber\\
    &\quad - \frac{1}{2}\braket{\varphi_j|U^\dagger(\bm{\theta}^*) A U(\bm{\theta}^*) |\varphi_j} \\
    \mathrm{Im}(\braket{\varphi_i|U^\dagger(\bm{\theta}^*) A U(\bm{\theta}^*) |\varphi_j}) &=
    \braket{+^y_{ij}|U^\dagger(\bm{\theta}^*) A U(\bm{\theta}^*) |+^y_{ij}} \nonumber\\
    &\quad - \frac{1}{2}\braket{\varphi_i|U^\dagger(\bm{\theta}^*) A U(\bm{\theta}^*) |\varphi_i} \nonumber\\
    &\quad - \frac{1}{2}\braket{\varphi_j|U^\dagger(\bm{\theta}^*) A U(\bm{\theta}^*) |\varphi_j}
\end{align}
where $\ket{+^x_{ij}} = (\ket{\varphi_i} + \ket{\varphi_j})/\sqrt{2}$ and $\ket{+^y_{ij}} = (\ket{\varphi_i} + i\ket{\varphi_j})/\sqrt{2}$.
Recall that in practice the input states are chosen from simple states such as the computational basis, and therefore we assume that the superpositions like $\ket{+^x_{ij}}$ can easily be prepared.
Each term of the above equation are measured separately on the NISQ device and then are summed up on a classical computer.

%% file: 60_experiment.tex
\section{Numerical Simulation}\label{sec:numerical_simulation}

\begin{figure}[H]
    \centering
    \includegraphics[width=1.0\linewidth]{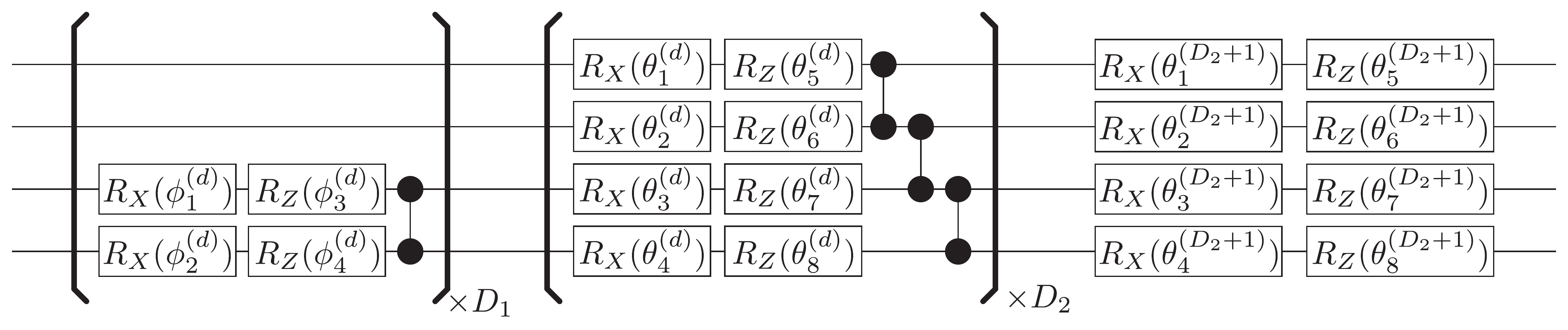}
    \caption{
        Variational quantum circuit used in the simulations of \cref{sec:numerical_simulation}.
        These parameters $\bm{\phi}, \bm{\theta}$ are optimized to to minimize $\loss$. $D_1$ and $D_2$ denote the number of repetition of a circuit in each bracket.
        Note that, in the explanation of weighted SSVQE, $\bm{\theta}$ denotes $\{\bm{\phi}, \bm{\theta}\}$ in this figure.
    }
    \label{fig:gates}
\end{figure}

Here we numerically simulate our algorithms with 4-qubit Hamiltonians.
\Cref{fig:gates} shows the variational ansatz used in the simulations.
We chose the input states as $\left\{\ket{\varphi_j}\right\} = \left\{\ket{0000}, \ket{0001}, \ket{0010}, \ket{0011} \right\}$.
The depth $D_1$ is set to $D_1 = 2$ for all of them.
$D_2$ is set to $D_2=6$ for the SSVQE and the weighted SSVQE for finding the $k$-th excited states, and $D_2=8$ for the weighted SSVQE for finding the all excited states up to the $k$-th.
The initial values of the parameters were randomly sampled from a uniform distribution $[0, 2\pi)$.
For each simulation, the optimization was run for 10 times starting from different initial values.
The results shown in the following sections are the ones which achieved the lowest value of the cost function among those 10 results.
We used the BFGS method~\cite{Nocedal2006} implemented in the SciPy library \cite{SciPy} for the optimization of the parameters.

\subsection{Transverse Ising model}

First, we demonstrate our idea with a Hamiltonian of the fully connected transverse Ising model:
\begin{equation}
    H = \sum_{i=1}^N a_i X_i + \sum_{i=1}^N \sum_{j=1}^{i-1} J_{ij} Z_i Z_j,
\end{equation}
with $N = 4$.
The coefficients $a_i$ and $J_{ij}$ are sampled randomly from a uniform distribution on $[0, 1)$.
In this subsection, we use one Hamiltonian with the same coefficients as an example.
All experiments were conducted on the case of $k = 3$.

\subsubsection{SSVQE}
\begin{figure}[H]
    \centering
    \includegraphics[width=0.8\linewidth]{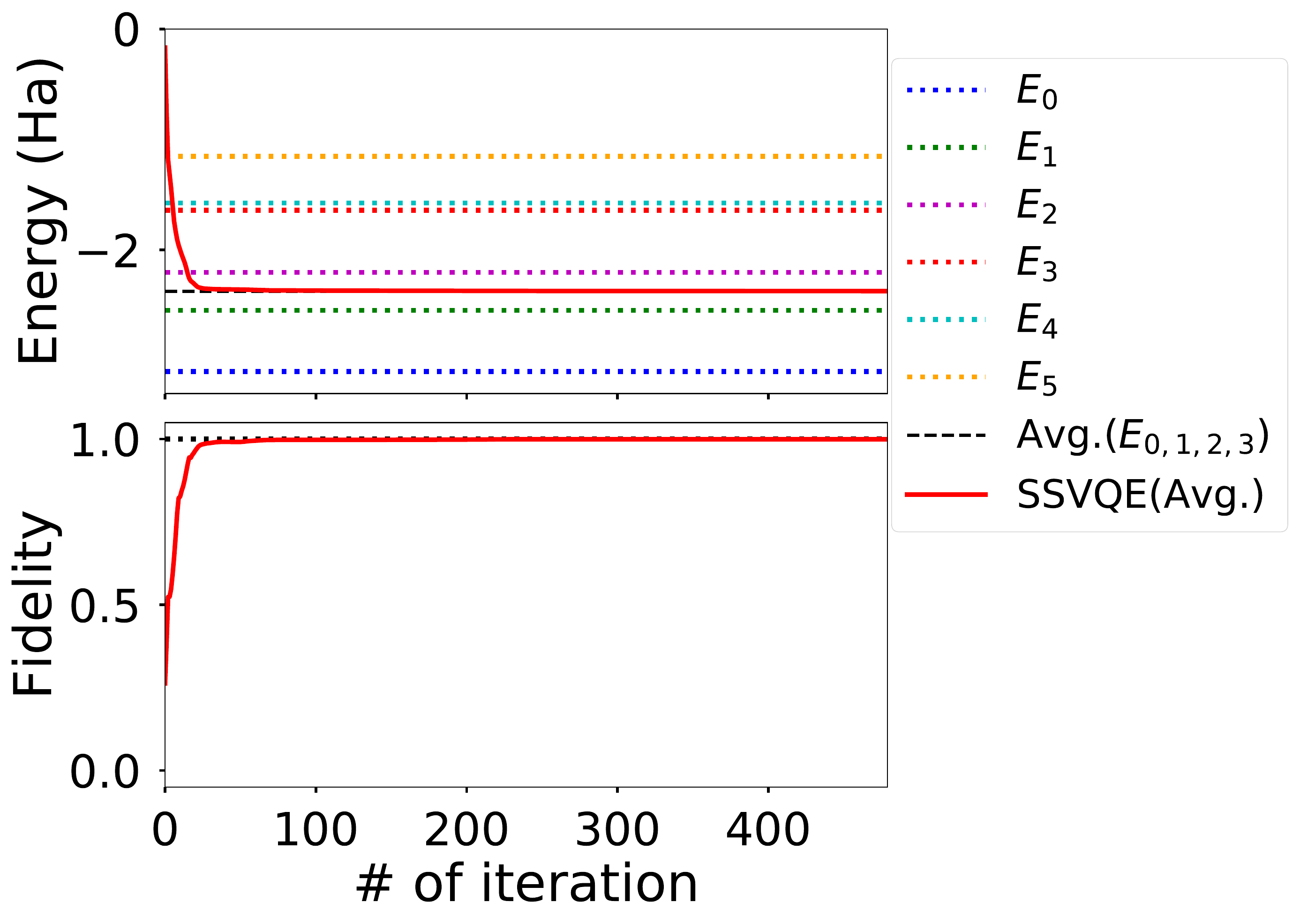}
    \caption{
        Step 1 of the SSVQE to find third excited state of a transverse Ising model. (black dashed line) $\mathrm{Avg.}(E_{0,1,2,3})$ = $\frac{1}{4}\sum_{k=0}^3 E_k$ , which is the globally optimal value of $\loss_1/4$ in this case.
        (red solid lines) The evolution of $\loss_1/4$ and the fidelity (see the main text for the definition) during the optimization process.
    }
    \label{fig:SS3_ising_step1}
\end{figure}

The SSVQE can find the $k$-th excited state with only two optimization procedures.
\Cref{fig:SS3_ising_step1} shows the first optimization process of $\bm{\theta}$ to minimize $\loss_1(\bm{\theta})$.
In \cref{fig:SS3_ising_step1}, the fidelity is defined by the overlap between the space spanned by $\left\{\ket{E_j}\right\}_{j=0}^{3}$ and the output of the quantum circuit $\left\{U(\bm{\theta})\ket{\varphi_j}\right\}_{j=0}^3$, namely, $\frac{1}{4}\sum_{i=0}^{3}\sum_{j=0}^{3}\left|\braket{E_i|U(\bm{\theta})|\varphi_j}\right|^2$.
We see that, as the cost function gets close to its global minimum, the fidelity approaches unity as expected.

\begin{figure}[H]
    \centering
    \includegraphics[width=0.8\linewidth]{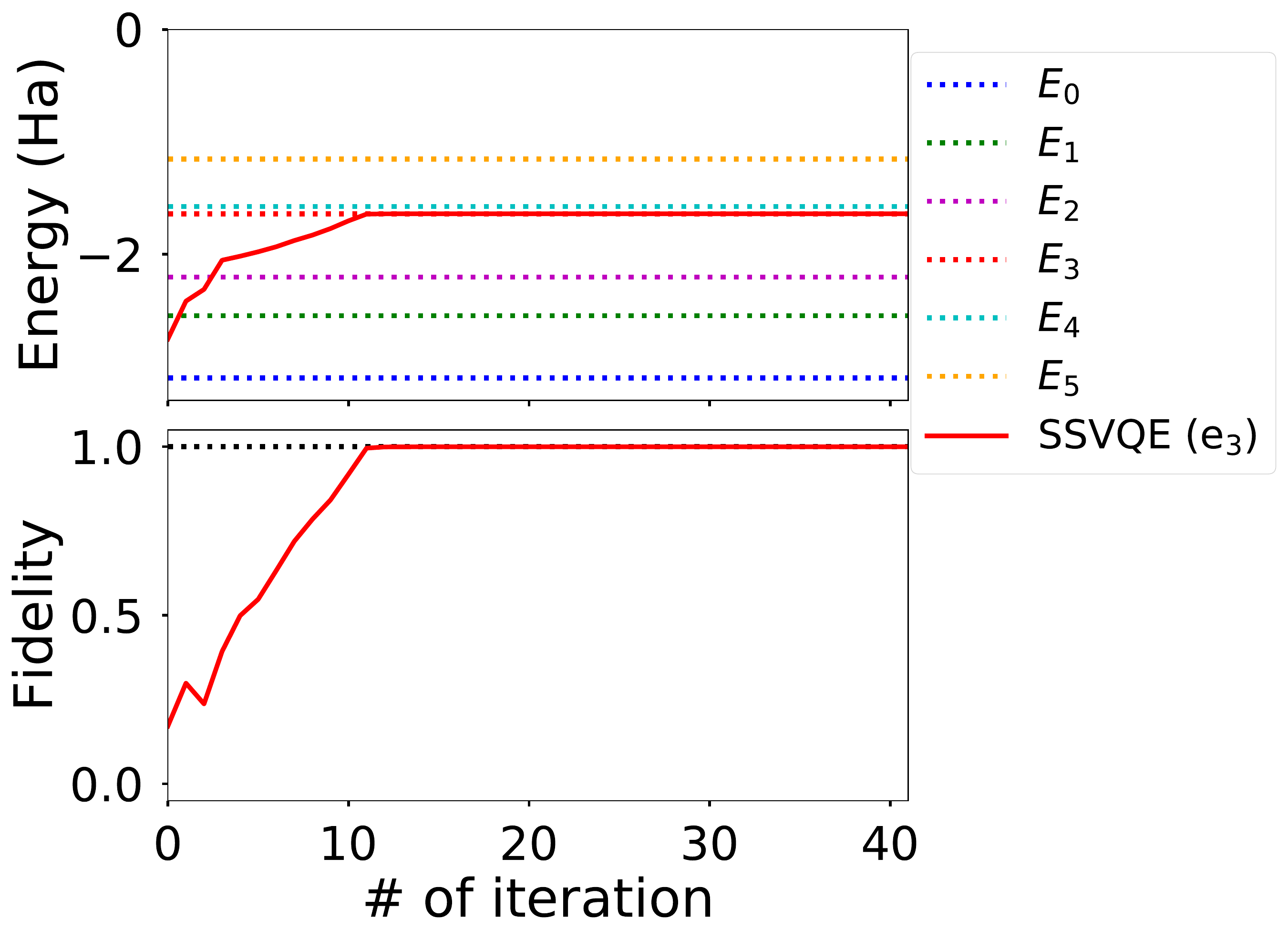}
    \caption{
        Step 2 of the SSVQE to find the third excited state of a transverse Ising model.
        (red solid lines) The evolution of $\loss_2$ and the fidelity (see the main text for the definition) during the optimization process.
    }
    \label{fig:SS3_ising_step2}
\end{figure}

\Cref{fig:SS3_ising_step2} shows the second process of optimizing $\bm{\phi}$ to minimize $\loss_2(\bm{\phi})$.
Here the fidelity is defined by $\left|\braket{E_3|U(\bm{\theta^*})V(\bm{\phi})|\varphi_3}\right|^2$.
One can see the subspace-search approach works well from \cref{fig:SS3_ising_step2}.


\subsubsection{Weighted SSVQE for finding the $k$-th excited state}
\begin{figure}[H]
    \centering
    \includegraphics[width=0.8\linewidth]{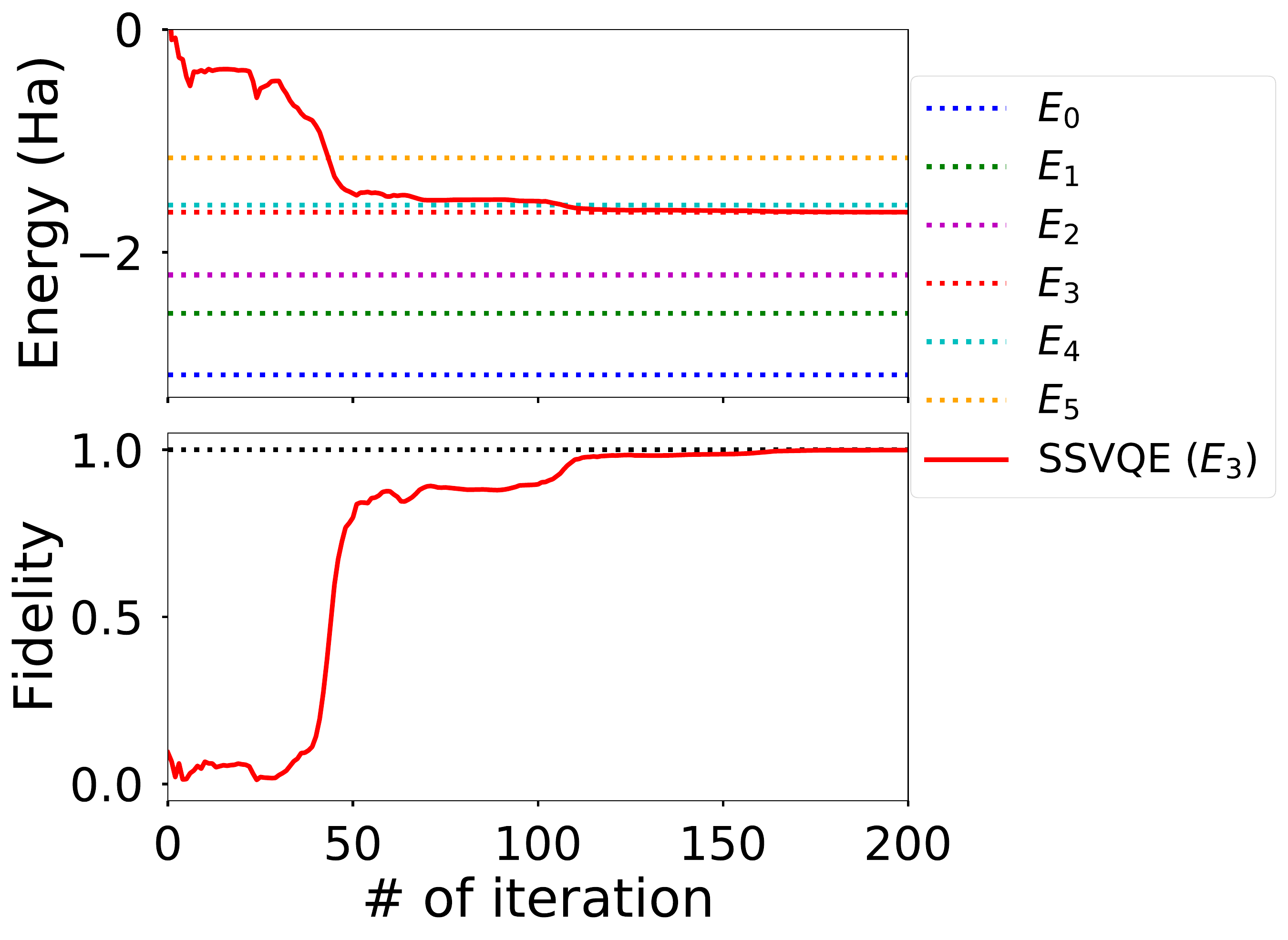}
    \caption{
        The weighted SSVQE to find the third excited state of a transverse Ising model.
        In the energy diagram, SSVQE($E_3$) (red solid line) is $\braket{\varphi_3|U^\dagger(\bm{\theta})HU(\bm{\theta})|\varphi_3}$ at each iteration.
    }
    \label{fig:E3_ising}
\end{figure}

The method described in \cref{sec:kth-WSSVQE} can find the $k$-th excited state by only one optimization sequence.
Here, we chose $w = 0.5$ as the weight.
\Cref{fig:E3_ising} shows the optimization process of $\bm{\theta}$ to minimize $\loss_w(\bm{\theta})$.
Here the fidelity is defined by $\left|\braket{E_3|U(\bm{\theta})|\varphi_3}\right|^2$.
In this case, too, the algorithm succeeds in finding the third excited state of the Hamiltonian.
However, the number of iterations to the convergence is larger than the number of overall iterations of the simple SSVQE.
It might be attributed to the more complicated landscape existing in the cost function.

\subsubsection{Weighted SSVQE}
\begin{figure}[H]
    \centering
    \includegraphics[width=0.8\linewidth]{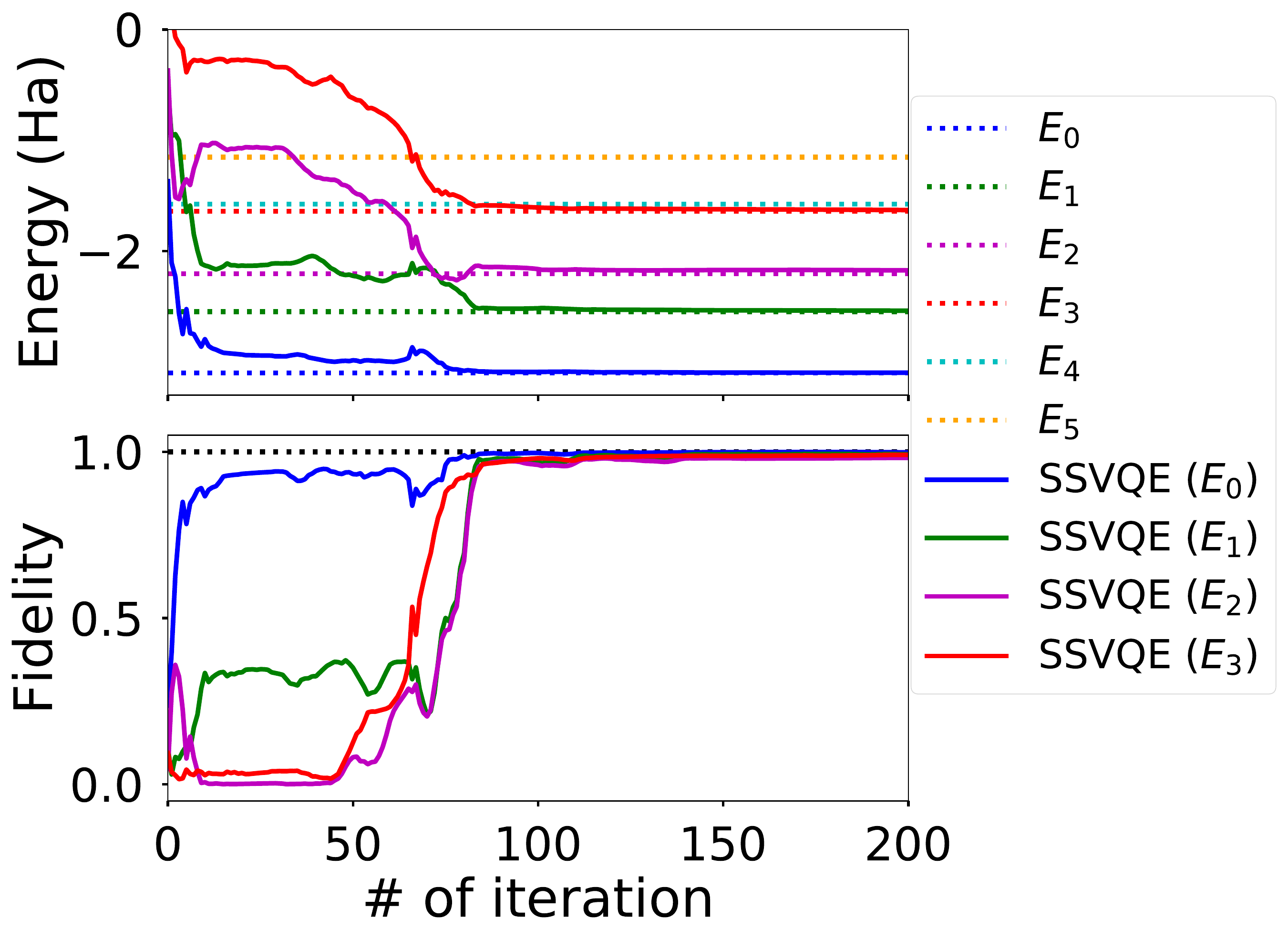}
    \caption{
        The weighted SSVQE to find excited states of a transverse Ising model up to the third. In the energy diagram, SSVQE($E_k$) (solid lines) are $\braket{\varphi_k|U^\dagger(\bm{\theta})HU(\bm{\theta})|\varphi_k}$ for each $k$ at each iteration.
    }
    \label{fig:E03_ising}
\end{figure}

The weighted SSVQE described in \cref{sec:WSSVQE} can find $0, 1, \cdots, k$-th excited states all at once.
Here, we chose $\bm{w} = (4, 3, 2, 1)$ as the weight vector.
\Cref{fig:E03_ising} shows the optimization process of $\bm{\theta}$ to minimize $\loss_{\bm{w}}(\bm{\theta})$.
From \cref{fig:E03_ising}, one can see that this approach can actually find the desired excited states all at once.
The number of iterations to the convergence is almost equivalent to the one presented in the previous section.

\subsection{Helium hydride}
Next, we apply our idea for the molecular Hamiltonians of $\text{HeH}$ with a fixed distance between two atoms.
Our ansatz (\cref{fig:gates}) does not consider the conservation of the number of electrons, and therefore, the calculated excited states can have the different number of them.
The molecular Hamiltonians are calculated with OpenFermion and OpenFermion-Psi4~\cite{McClean2017, Parrish2017}.
We used the STO-3G minimal basis set, and therefore, obtained the 4-qubit Hamiltonian.
We calculated the Hamiltonians at the 24 different bond lengths and performed the VQE at each point.
In the weighted SSVQE simulation, we used the same weights as in the previous section.

\begin{figure}[H]
    \centering
    \includegraphics[width=0.7\linewidth]{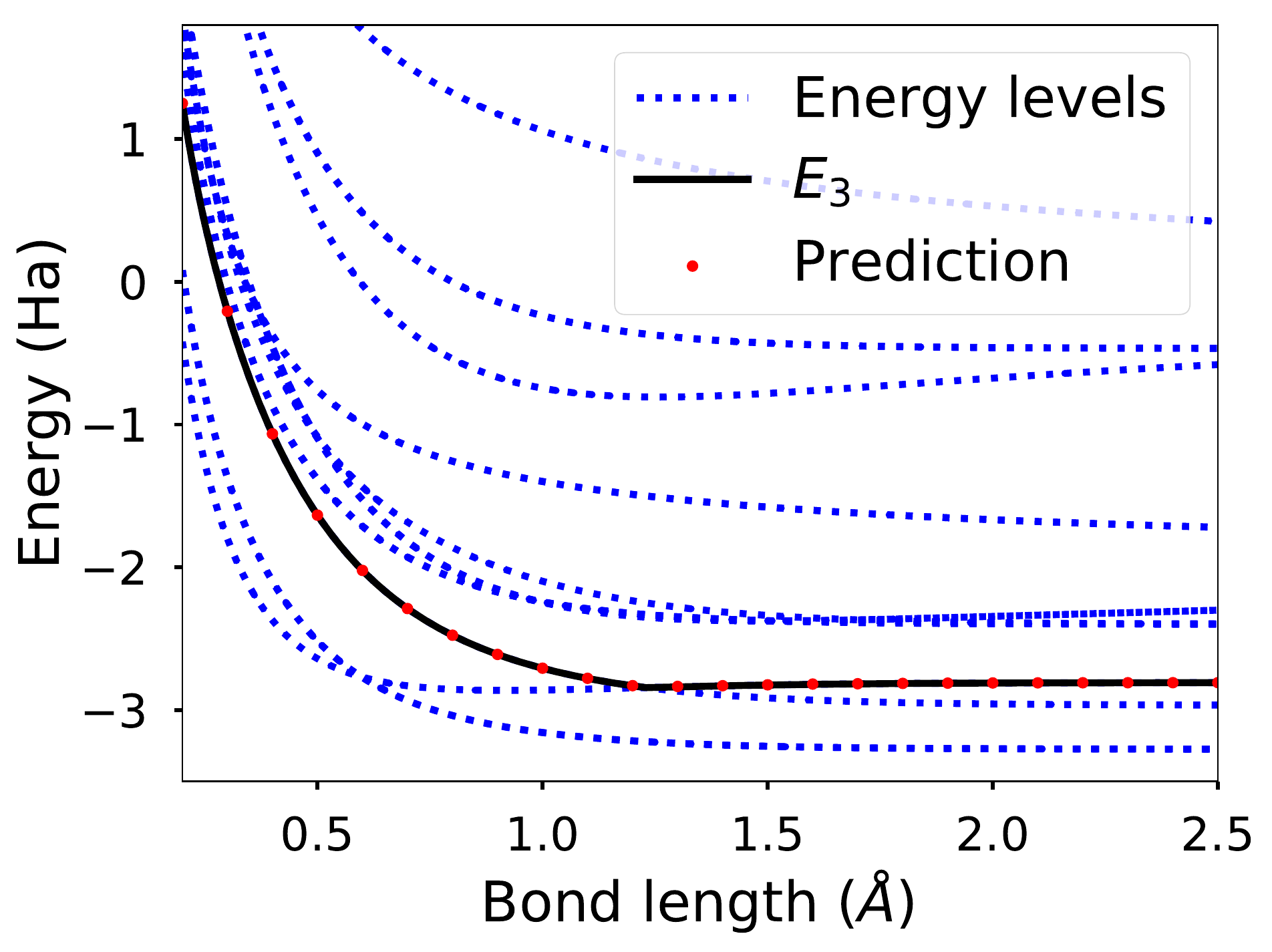}
    \caption{\label{fig:HeH_SSVQE}
        The energy levels of the Hamiltonian of $\mathrm{HeH}$ and the calculated energy of the third excited state using the SSVQE.
    }
    \label{fig:hoge}
\end{figure}

\begin{figure}[H]
    \centering
    \includegraphics[width=0.7\linewidth]{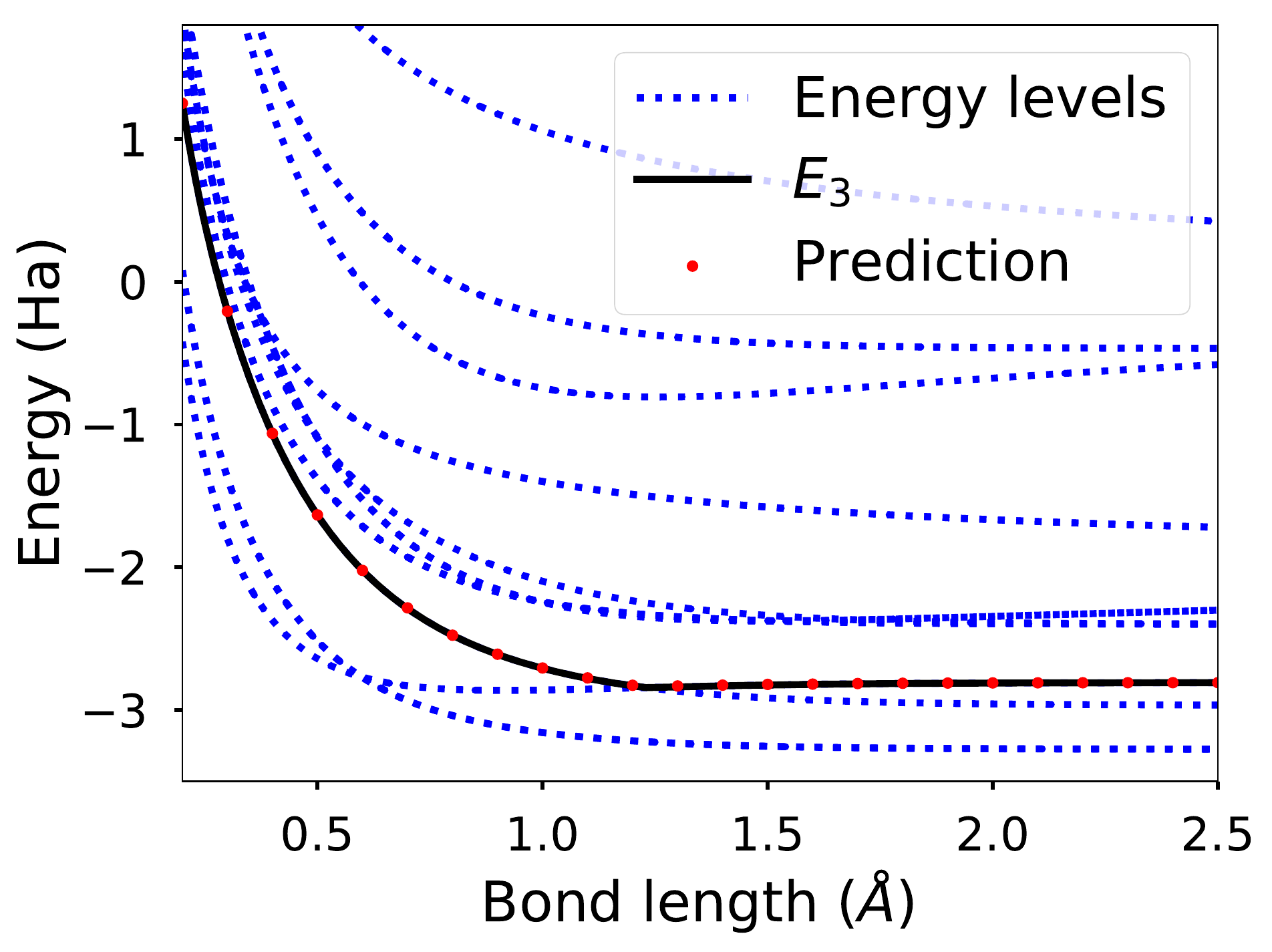}
    \caption{\label{fig:}
        The energy levels of the Hamiltonian of $\mathrm{HeH}$ and the predicted energy of the third excited state using the weighted SSVQE.
    }
    \label{fig:HeH_WSSVQE_only_k}
\end{figure}

The result of SSVQE is shown in \cref{fig:HeH_SSVQE} and one from using weighted SSVQE for finding the $k$-th excited state is shown in \cref{fig:HeH_WSSVQE_only_k}.
Both of the results agree nicely with the exact values of the third excited state at each bond length.

\begin{figure}[H]
    \centering
    \begin{minipage}{0.49\linewidth}
        \centering
        \includegraphics[width=\linewidth]{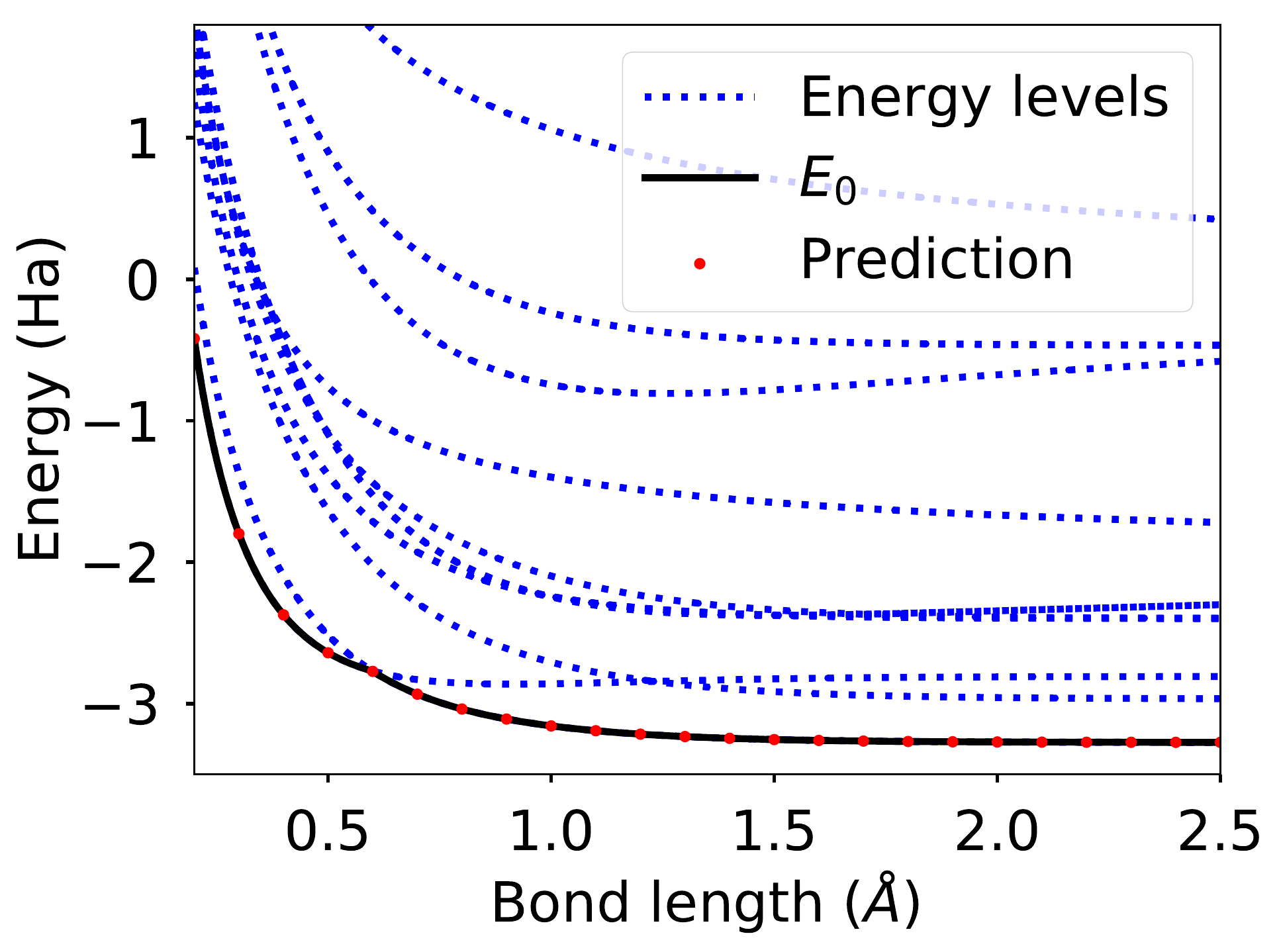}
        \subcaption{}\label{fig:HeH_WSSVQE_0}
    \end{minipage}
    \begin{minipage}{0.49\linewidth}
        \centering
        \includegraphics[width=\linewidth]{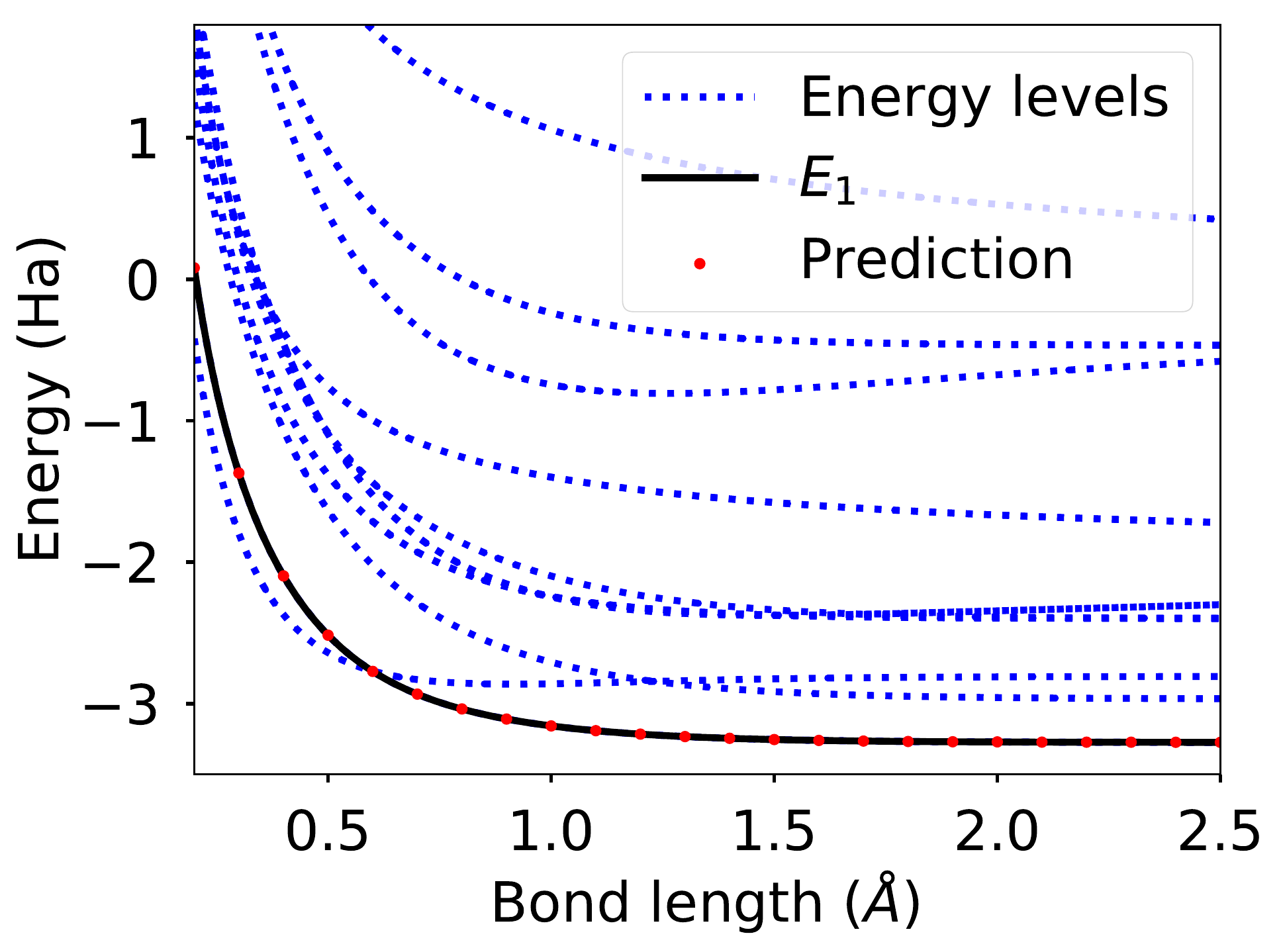}
        \subcaption{}\label{fig:HeH_WSSVQE_1}
    \end{minipage}\\
    \begin{minipage}{0.49\linewidth}
        \centering
        \includegraphics[width=\linewidth]{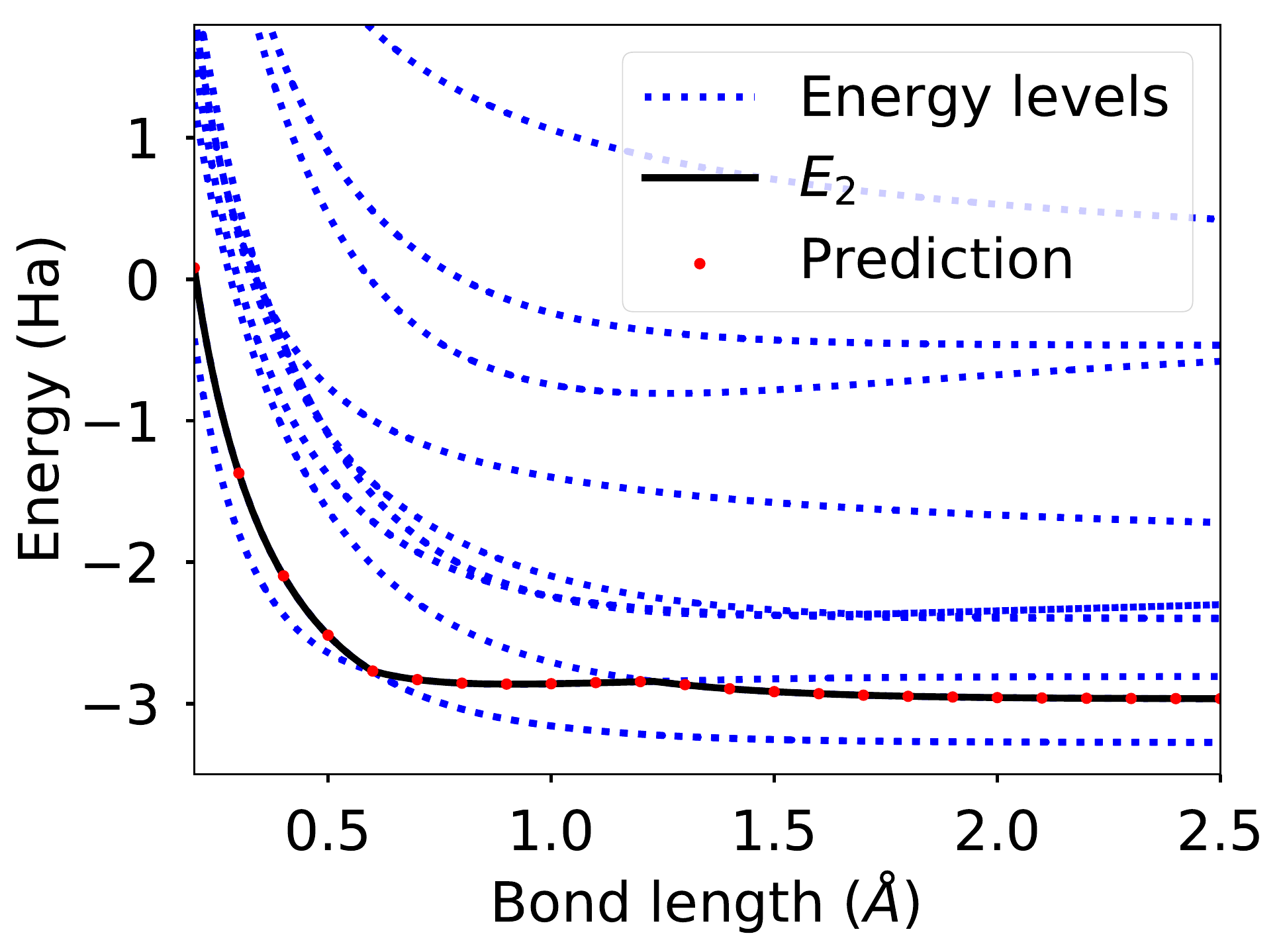}
        \subcaption{}\label{fig:HeH_WSSVQE_2}
    \end{minipage}
    \begin{minipage}{0.49\linewidth}
        \centering
        \includegraphics[width=\linewidth]{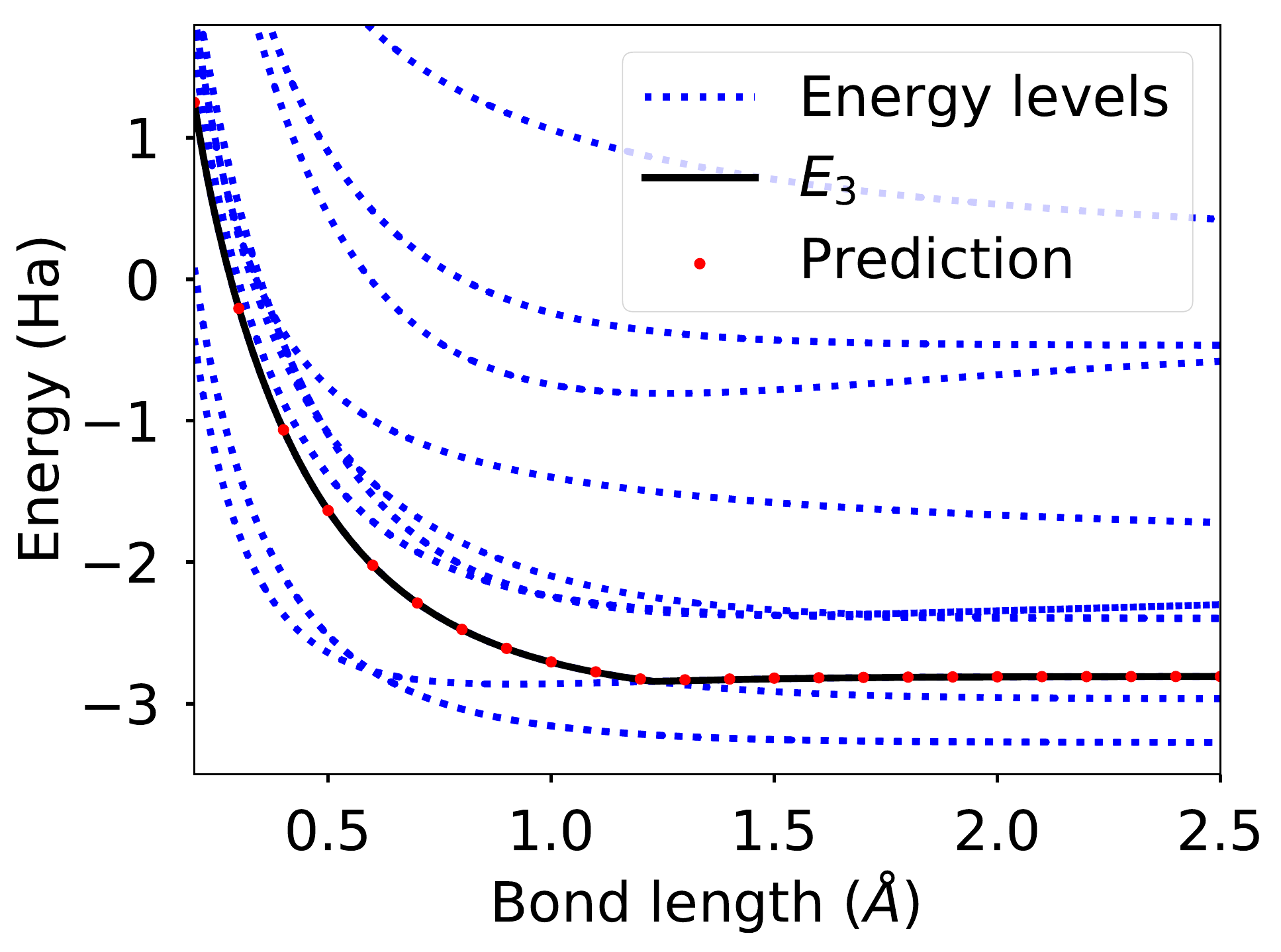}
        \subcaption{}\label{fig:HeH_WSSVQE_3}
    \end{minipage}
    \caption{
        The energy levels of the Hamiltonian of $\mathrm{HeH}$ and the predicted energy of the excited states up to the third using the weighted SSVQE.
        (a) 
        ground state
		(b)
        1st excited state
		(c)
        2nd excited state
		(d)
        3rd excited state
    }
    \label{fig:HeH_WSSVQE}
\end{figure}

Next, we used the weighted SSVQE described in \cref{sec:WSSVQE} to find all excited states up to the third.
The result is shown in Fig.~\ref{fig:HeH_WSSVQE}.
One can see that the energy eigenvalues are well approximated by the optimized output of the weighted SSVQE.

%% file: 80_conclusion.tex
\section{Conclusion}
In this work, we proposed efficient algorithms for finding excited states of a given Hamiltonian, extending the framework of the VQE.
The proposed method assures the orthogonality of the states at the input of the ansatz circuit.
Minimizing a carefully designed cost function by optimizing the parameters of the quantum circuit, we can map each of the orthogonal states onto one of the energy eigenstates.
Our algorithms require us to run, in principle, optimization only once or twice, and find one or more arbitrary excited states.
We believe that this work greatly extends the practicability of the VQE for finding the excited states.

%% file: 85_ack.tex
\section*{Acknowledgements}\label{sec:acknowledgements}
This work was supported by QunaSys Inc.\footnote{\url{https://qunasys.com/}}